\documentclass[preprintnumbers,superscriptaddress,amsmath,amssymb,prd,preprint,nofootinbib]{revtex4-2}
\usepackage{amssymb}
\usepackage{mathrsfs}
\usepackage{booktabs}
\usepackage{amsthm}
\usepackage{graphicx}
\usepackage{amsfonts}
\usepackage[colorlinks,linkcolor=red,anchorcolor=blue,citecolor=green]{hyperref}
\usepackage{subfigure}
\usepackage{float}
\usepackage{enumerate}
\usepackage{multirow}

\begin{document}	
\title{Coordinate Independence of the Schwarzschild Black Hole Accretion Vlasov Gas Model}
\author{Ping Li}
\email[]{Lip57120@huas.edu.cn}
\affiliation{College of Mathematics and Physics, Hunan University of Arts and Sciences, 3150 Dongting Dadao, Changde City, Hunan Province 415000, China}
\affiliation{Hunan Province Key Laboratory Integration of Photoelectric Information Integration and Optical Manufacturing Technology, 3150 Dongting Dadao, Changde City, Hunan Province 415000, China}
\author{Jun Cheng}
\email[]{chengjun@huas.edu.cn}
\affiliation{College of Mathematics and Physics, Hunan University of Arts and Sciences, 3150 Dongting Dadao, Changde City, Hunan Province 415000, China}
\affiliation{Hunan Province Key Laboratory Integration of Photoelectric Information Integration and Optical Manufacturing Technology, 3150 Dongting Dadao, Changde City, Hunan Province 415000, China}

\author{Jiang-he Yang}
\email[]{yjianghe@163.com}
\affiliation{College of Mathematics and Physics, Hunan University of Arts and Sciences, 3150 Dongting Dadao, Changde City, Hunan Province 415000, China}
\affiliation{Center for Astrophysics, Guangzhou University, 230 West Ring Road, Guangzhou, Guangdong Province 510006, China}

\begin{abstract}
This paper presents a detailed study of the coordinate dependence of Vlasov gas accretion onto a Schwarzschild black hole. Asymptotic results at infinity and near
horizon are obtained via Taylor expansions for three different statistical distributions within the framework of the most general stationary spherically symmetric spacetime. Our findings demonstrate that the particle number density, energy density, radial and tangential pressures, and accretion rates are independent of the coordinate choice, even though individual components such as the particle current density and the stress-energy tensor explicitly depend on the coordinate system. Consequently, the accretion theory can be formulated without reference to any particular coordinate system. We also show that the mean energy of the accreted particles is $m_0+k_BT$, lower than the mean energy $m_0+\frac{3}{2}k_BT$ of the Maxwell-Boltzmann system in the classical limit. And the specific entropy of the accreted particles is lower than the global average by $\frac{3}{2}k_B$. This is because particles of lower energy are more easily accreted, while particles of higher energy are more readily scattered. We also present numerical results at finite radii for the relevant physical quantities.

\end{abstract}
\maketitle

\section{Introduction}
The problem of stationary, spherically symmetric accretion of matter onto a black hole has a long history, dating back to the classical work of Bondi~\cite{Bondi1952} and Michel~\cite{Michel1972} in the hydrodynamic regime. In these seminal papers, the gas was modeled as a perfect fluid, and explicit solutions for the mass accretion rate were obtained under the assumption of a polytropic equation of state. However, in high-temperature, low-density astrophysical environments such as those observed around the supermassive black holes at the centers of the Galaxy (Sgr A*) and M87~\cite{EHT2019, EHT2022}, the mean free path of particles can greatly exceed the characteristic system scale, rendering collisional effects negligible. In such collisionless or weakly collisional plasmas, the hydrodynamic description breaks down, and a kinetic approach based on the Vlasov equation becomes essential. The formal structure of relativistic kinetic theory on curved spacetimes has been developed over decades, with foundational contributions by Synge~\cite{Synge1934}, Ehlers~\cite{Ehlers1971}, and more recently by Sarbach and Zannias~\cite{Sarbach2013, Sarbach2014, Acuna2022}, who provided a rigorous geometric framework for the formulation of the Vlasov equation on the tangent and cotangent bundles. For a comprehensive review of the Einstein–Vlasov system, see Andr\'easson~\cite{Andreasson2011}.

In 2017, Rioseco and Sarbach provided the first systematic study of the accretion of a relativistic, collisionless kinetic gas (the Vlasov gas) onto a Schwarzschild black hole~\cite{Rioseco2017a, Rioseco2017b}. Working within the Hamiltonian formulation on the cotangent bundle, they solved the relativistic Liouville equation, derived the most general solution for the distribution function in terms of action-angle variables, and computed the associated observables including the particle current density and stress–energy–momentum tensor. In the steady-state, spherically symmetric case with an equilibrium distribution at infinity, they found that in the low-temperature limit the tangential pressure at the horizon is about an order of magnitude larger than the radial one, explicitly demonstrating that a collisionless gas behaves very differently from an isotropic perfect fluid. This provided a partial explanation for the known fact that the kinetic accretion rate is much lower than in the hydrodynamic Bondi–Michel case~\cite{Bondi1952, Michel1972}. Moreover, they established the asymptotic stability of the steady-state spherical flows by proving pointwise convergence results for a large class of initial conditions~\cite{Rioseco2017b}. The accretion of Vlasov gas onto a Schwarzschild black hole was also studied from the perspective of geodesic motion and analytical solutions for timelike and null geodesics by Cie\'slik and Mach~\cite{Cieslik2022, Cieslik2023}.

The Rioseco–Sarbach model has since been extended to a variety of spherically symmetric black hole spacetimes. Mach and Odrzywo{\l}ek derived exact axially symmetric solutions for accretion onto a moving Schwarzschild black hole~\cite{Mach2021a, Mach2021b}, providing a relativistic counterpart of the Bondi–Hoyle–Lyttleton problem, and further extended this work to the low-temperature limit and numerical aspects~\cite{Mach2022}. Cie\'slik and Mach generalized the analysis to Reissner–Nordstr\"om black holes~\cite{Cieslik2020}, showing that the mass accretion rate decreases with increasing black hole charge. Li, Yang, and Xu later studied the accretion of a degenerate Fermi gas using Fermi–Dirac statistics~\cite{Li2025}, demonstrating the effect of quantum statistics on the accretion process in a charged background. Liu and Zhang provided a complete analytical treatment of two-component plasma Vlasov gas accretion onto a Reissner–Nordstr\"om black hole~\cite{Liu2026}, addressing the electromagnetic interactions between charged particles. Gamboa et al. considered a generalization of the Bondi-type model where the boundary conditions are specified at a sphere of finite radius instead of infinity~\cite{Gamboa2021}, highlighting the role of angular momentum in the accretion flow. The model was further extended to regular black holes, such as the Bardeen black hole~\cite{Liao2022}, and to Schwarzschild-like black holes in modified gravity theories such as bumblebee gravity~\cite{Cai2023}. Momennia and Sarbach recently provided a general framework for spherical accretion of a collisionless kinetic gas into generic static, asymptotically flat black hole spacetimes~\cite{Momennia2025}. Zhang and Jiang investigated Vlasov gas accretion onto a black hole in the Kalb–Ramond field~\cite{Zhang2025}. In parallel, axisymmetric stationary collisionless gas configurations surrounding Schwarzschild black holes were analyzed by Gabarrete and Sarbach~\cite{Gabarrete2022, Gabarrete2023}, and Shapiro studied spikes and accretion of unbound collisionless matter around black holes~\cite{Shapiro2023}.

The extension of the Vlasov gas accretion model to rotating black hole spacetimes has proven to be considerably more challenging due to the loss of spherical symmetry and the complexity of the Kerr geodesic structure. Mach, Momennia, and Sarbach established a complete accretion theory for a Vlasov gas in Kerr spacetime~\cite{Mach2026a, Mach2026b}, providing a rigorous phase-space characterization of absorption and scattering orbits and deriving analytic slow-rotation approximations that accurately describe the accretion process even for rapidly rotating black holes. Li, Liu, and Zhai also provided integral expressions for the particle current density and stress–energy tensor in full (3+1)-dimensional Kerr spacetime and numerically computed the accretion rate per unit solid angle~\cite{Li2023}. Liu extended the analysis to Kerr–Newman spacetime and studied plasma Vlasov gas accretion under a weak electromagnetic coupling approximation~\cite{Liu2025}. Finite collisionless accretion disks in Kerr spacetime were constructed by Khan and Mach~\cite{Khan2025}, and stable collisionless tori around Kerr black holes were investigated by Luepker, Yuan, and Chen~\cite{Luepker2025}. The relaxation of macroscopic variables to a stationary state around a Kerr black hole due to phase-space mixing was studied by Rioseco and Sarbach~\cite{Rioseco2018, Rioseco2024}. The general properties of timelike and null geodesics in the Kerr spacetime, which are essential for the phase-space analysis of kinetic accretion, have been studied by Cie\'slik, Hackmann, and Mach~\cite{Cieslik2023} and by Tejeda, Taylor, and Miller~\cite{Tejeda2013}; a recent analysis in horizon-penetrating coordinates was given by Bakun et al.~\cite{Bakun2025}. Hydrodynamical accretion onto rotating black holes has also been studied numerically by Aguayo-Ortiz et al.~\cite{Aguayo2021}, providing a complementary perspective to the kinetic approach. The slow-rotation approximation for Kerr black holes, which is astrophysically relevant, was further analyzed by Pani, Berti, and Gualtieri in the context of gravitational perturbations~\cite{Pani2013}. An exact hydrodynamical solution for Bondi-type accretion onto a Kerr black hole was obtained by Petrich, Shapiro, and Teukolsky for an ultra-hard equation of state~\cite{Petrich1988}.

Monte Carlo methods have been developed for stationary solutions of general-relativistic Vlasov systems~\cite{Mach2023, Cieslik2024}, providing an alternative numerical approach to complement analytical studies. These methods are particularly useful for handling complex phase-space geometries and for validating analytical approximations in regimes where direct numerical integration becomes challenging.

Despite these advances, two fundamental aspects of the Vlasov gas accretion theory have not been fully clarified. First, the coordinate dependence of the formalism has not been systematically investigated. While individual components of the particle current density and the stress–energy tensor explicitly depend on the choice of coordinate system, physical observables such as the particle number density, energy density, and pressures should be coordinate-independent. Yet, previous works have typically been formulated in specific coordinate systems---such as Eddington–Finkelstein or Painlev\'e–Gullstrand coordinates---and a general proof of the invariance of all relevant physical quantities has been lacking. Second, the thermodynamic properties of the particles that are actually captured by the black hole---as opposed to those that are scattered back to infinity---have not been examined in detail. In particular, it remains unclear how the angular momentum filtering mechanism, which distinguishes absorbed from scattered trajectories, affects the mean energy and entropy of the accreted particles, and how this compares with the classical Maxwell–Boltzmann expectation.

In this paper, we address both of these open questions. We perform our calculations within the framework of the most general stationary spherically symmetric spacetime, which allows us to consider different coordinate representations simultaneously---including orthogonal Schwarzschild coordinates, Eddington–Finkelstein coordinates, and Painlev\'e–Gullstrand-type coordinates---and to demonstrate explicitly that the accretion theory can be formulated without reference to any particular coordinate system. We show that although the components \(J_\mu\) and \(T_{\mu\nu}\) depend on the coordinate choice, the eigenvalues of the mixed stress–energy tensor \(T^\mu{}_\nu\) (giving the energy density \(\rho\), radial pressure \(p_{\mathrm{rad}}\), and tangential pressure \(p_{\mathrm{tan}}\)) and the particle number density \(n = \sqrt{-g_{\mu\nu}J^\mu J^\nu}\) are coordinate-invariant scalars. Furthermore, we prove that the particle and energy accretion rates \(\dot{n}\) and \(\dot{E}\), defined via the conserved currents associated with the Killing symmetries, are also coordinate-independent, with the identity \(a^2b^2+h^2=1\) ensuring that the formulas take the simple form \(-4\pi r^2 J^r\) and \(-4\pi r^2 T^r{}_t\) in all coordinates considered. We also examine the thermodynamic properties of the accreted particles for three different quantum statistics: Fermi–Dirac, Maxwell–J\"uttner, and Bose–Einstein. In the classical limit, we find that the mean energy of the particles that fall into the black hole is \(m_0 + k_B T\), which is lower than the mean energy \(m_0 + \frac{3}{2} k_B T\) of the Maxwell–J\"uttner gas at infinity. This reduction arises because high-energy particles tend to carry large angular momenta, making them more likely to be scattered by the centrifugal barrier. Consequently, the specific entropy of the accreted particles is lower than the global average by \(\frac{3}{2}k_B\). Our results demonstrate that the accretion theory can be formulated in a coordinate-invariant manner, and that the angular momentum selection effect has important consequences for the thermodynamic evolution of the black hole–gas system.

The paper is organized as follows. In Sec.~II, we present the general metric for a stationary spherically symmetric spacetime and derive the equations of motion for test particles. In Sec.~III, we construct the action-angle variables that trivialize the Liouville equation. In Sec.~IV, we introduce the relativistic equilibrium distributions for three different quantum statistics: Fermi–Dirac, Maxwell–J\"uttner, and Bose–Einstein. In Sec.~V, we derive the integral expressions for the particle current density and the stress–energy–momentum tensor. In Sec.~VI, we discuss the limits of integration and the classification of absorption and scattering orbits. In Sec.~VII, we identify the coordinate-independent physical quantities and discuss the diagonalization of the stress–energy tensor. In Secs.~VIII and IX, we analyze the asymptotic behavior at infinity and near the horizon, respectively. In Sec.~X, we specialize to the Maxwell–J\"uttner distribution and derive the accretion rates of thermodynamic quantities. In Sec.~XI, we present numerical results for finite radii. Finally, Sec.~XII contains our conclusions and discussion.

\section{Equations of motion for test particles in a general spherically symmetric stationary spacetime}

In this paper, we consider a general spherically symmetric stationary spacetime with the metric 
\begin{equation}
ds^2 = -b^2(r) dt^2 - 2h(r) dt dr + a^2(r) dr^2 + r^2 d\theta^2 + r^2 \sin^2\theta d\varphi^2,
\label{eq:metric}
\end{equation}
where \(b(r)\), \(h(r)\) and \(a(r)\) are arbitrary functions of the radial coordinate \(r\). This form contains a non-diagonal component \(g_{tr} = -h(r)\); therefore \(\partial_t\) is a timelike Killing vector, but because \(g_{tr} \neq 0\) this Killing vector is not orthogonal to the hypersurfaces \(t = \text{constant}\), so the spacetime is stationary but not static. The determinant of the metric is \(\det(g) = -r^4 \sin^2\theta (a^2 b^2 + h^2)\), and the inverse metric can be written in matrix form as
\begin{equation}
g^{\mu\nu} =
\begin{pmatrix}
-\dfrac{a^2}{a^2 b^2 + h^2} & -\dfrac{h}{a^2 b^2 + h^2} & 0 & 0 \\[6pt]
-\dfrac{h}{a^2 b^2 + h^2} & \dfrac{b^2}{a^2 b^2 + h^2} & 0 & 0 \\[6pt]
0 & 0 & \dfrac{1}{r^2} & 0 \\[6pt]
0 & 0 & 0 & \dfrac{1}{r^2 \sin^2\theta}
\end{pmatrix}.
\label{eq:inverse_metric}
\end{equation}

The metric form (\ref{eq:metric}) allows us to study the influence of different coordinate choices on the accretion theory. For instance, in the case of an orthogonal spacetime, the Schwarzschild black hole is described by the standard form
\[
b^{2}(r) = a^{-2}(r) = 1 - \frac{2M}{r}, \qquad h(r) = 0.
\]
Alternatively, one may adopt the Eddington-Finkelstein coordinates. Introducing the transformation
\[
u = t + \int^{r} \left( \frac{1}{b^{2}(r')} - \eta(r') \right) dr',
\]
the metric (\ref{eq:metric}) can be transformed into
\[
ds^{2} = -b^{2} du^{2} + 2(1 - b^{2}\eta) du dr + \eta(2 - b^{2}\eta) dr^{2} + r^{2}(d\theta^{2} + \sin^{2}\theta d\phi^{2}),
\]
where we now have \(b^{2} = 1 - 2M/r\), \(a^{2} = \eta(2 - b^{2}\eta)\), and \(h = 1 - b^{2}\eta\). In this work, we focus on three specific cases: the orthogonal (static) Schwarzschild coordinates (\(h=0\)), the Eddington-Finkelstein form with  \(h=1\)  (or \(\eta = 0\)), and the case \(h=2M/r\)  (or \(\eta = 1\)). Note that the case \(h=2M/r\) has been extensively discussed in the literature. In the Schwarzschild metric,  it can be shown that
\begin{equation}
    a^2b^2+h^2=1.
\end{equation}
Moreover, by employing the general metric (\ref{eq:metric}), our results can be readily extended to Reissner-Nordström black holes and even to asymptotically flat black holes in modified theories of gravity.

Consider a test particle of mass \(m\) moving in this background. Its dynamics is described by the Lagrangian
\begin{equation}
\mathcal{L} = \frac{1}{2} g_{\mu\nu} \frac{dx^\mu}{d\lambda} \frac{dx^\nu}{d\lambda},
\label{eq:Lagrangian}
\end{equation}
where \(\lambda\) is an affine parameter. Substituting the metric \eqref{eq:metric} into \eqref{eq:Lagrangian} gives
\begin{equation}
\mathcal{L} = -\frac{1}{2} b(r)^2 \dot{t}^2 - h(r) \dot{t} \dot{r} + \frac{1}{2} a(r)^2 \dot{r}^2 + \frac{1}{2} r^2 \dot{\theta}^2 + \frac{1}{2} r^2 \sin^2\theta \dot{\phi}^2,
\label{eq:Lagrangian_explicit}
\end{equation}
where the dot denotes differentiation with respect to \(\lambda\). From the Lagrangian we obtain the canonical momenta:
\begin{align}
p_t &= \frac{\partial \mathcal{L}}{\partial \dot{t}} = -b^2(r) \dot{t} - h(r) \dot{r}, \label{eq:pt}\\
p_r &= \frac{\partial \mathcal{L}}{\partial \dot{r}} = -h(r) \dot{t} + a(r)^2 \dot{r}, \label{eq:pr}\\
p_\theta &= \frac{\partial \mathcal{L}}{\partial \dot{\theta}} = r^2 \dot{\theta}, \label{eq:ptheta}\\
p_\phi &= \frac{\partial \mathcal{L}}{\partial \dot{\phi}} = r^2 \sin^2\theta \dot{\phi}. \label{eq:pphi}
\end{align}

Since the metric does not depend on \(t\) and \(\phi\), the corresponding canonical momenta are conserved. We define the energy \(E\) and the angular momentum \(l_z\) of the particle as
\begin{align}
E &\equiv -p_t = b^2(r) \dot{t} + h(r) \dot{r}, \label{eq:E}\\
l_z &\equiv p_\phi = r^2 \sin^2\theta \dot{\phi}. \label{eq:lz}
\end{align}
Spherical symmetry also guarantees the conservation of the total angular momentum squared \(l^2\). The Hamiltonian obtained via Legendre transformation is
\begin{equation}
H = p_\mu \dot{x}^\mu - \mathcal{L} = \frac{1}{2} g_{\mu\nu} \dot{x}^\mu \dot{x}^\nu,
\label{eq:Hamiltonian}
\end{equation}
and its conservation gives the rest mass constraint
\begin{equation}
m^2 = -2H = -g_{\mu\nu} \dot{x}^\mu \dot{x}^\nu.
\label{eq:mass_constraint}
\end{equation}

To obtain the radial equation of motion we need to solve for \(\dot{t}\) and \(\dot{r}\) from \eqref{eq:E} and \eqref{eq:pr}. Writing them in matrix form,
\begin{equation}
\begin{pmatrix}
b^2(r) & h(r) \\
h(r) & -a^2(r)
\end{pmatrix}
\begin{pmatrix}
\dot{t} \\
\dot{r}
\end{pmatrix}
=
\begin{pmatrix}
E \\
-p_r
\end{pmatrix},
\label{eq:velocity_matrix}
\end{equation}
we find
\begin{align}
\dot{t} &= \frac{a^2(r) E + h(r) p_r}{a^2(r) b^2(r) + h^2(r)}, \label{eq:tdot}\\
\dot{r} &= \frac{-h(r) E - b^2(r) p_r}{a^2(r) b^2(r) + h^2(r)}. \label{eq:rdot}
\end{align}

Substituting \eqref{eq:tdot} and \eqref{eq:rdot} into the mass constraint \eqref{eq:mass_constraint} and using \eqref{eq:lz} together with the expression for \(p_\theta\), we obtain after simplification an effective one‑dimensional equation for the radial motion. In terms of the conserved quantities,
\begin{equation}
\dot{r}^2 = \frac{1}{a^2(r) b^2(r) + h^2(r)} \left[ E^2 - m^2 b^2(r) - \frac{b^2(r) l^2}{r^2} \right],
\label{eq:rdot_squared}
\end{equation}
where \(l^2 = p_\theta^2 + p_\phi^2/\sin^2\theta\) is the square of the total angular momentum. To simplify later calculations we introduce the notation
\begin{equation}
R \equiv r^4 \dot{r}^2 = \frac{r^4}{a^2 b^2 + h^2} \left[ E^2 - b^2 \left( m^2 + \frac{l^2}{r^2} \right) \right].
\label{eq:R_definition}
\end{equation}
This is the effective kinetic energy for the radial motion. 

\section{Action-angle variables}

In this section we construct the action-angle variables for a neutral test particle moving in the general spherically symmetric stationary spacetime described by the metric \eqref{eq:metric}. The existence of cyclic coordinates \(t\) and \(\phi\) allows us to separate the Hamilton–Jacobi equation and to introduce conserved quantities that will serve as the new momenta. The action-angle variables then provide a convenient framework for solving the Vlasov equation.

The abbreviated action (or Hamilton's characteristic function) along a particle orbit is defined as
\begin{equation}
S= \int p_t dt + \int p_r dr + \int p_\theta d\theta + \int p_\phi d\phi,
\label{eq:action}
\end{equation}
where the integral is taken along the trajectory. Four conserved quantities are $p_t, p_\phi,l^2$ and \(m^2\). Substituting \(p_t = -E\) and \(p_\phi = l_z\) into the mass constraint gives
\begin{equation}
m^2 = -g^{tt}E^2 + 2g^{tr}E p_r - g^{rr}p_r^2 - \frac{l^2}{r^2}.
\label{eq:mass_constraint_momenta}
\end{equation}
Inserting the components of the inverse metric \eqref{eq:inverse_metric} and solving the resulting quadratic equation for \(p_r\) yields
\begin{equation}
p_r = -\frac{h E}{b^2} \pm \frac{\sqrt{a^2 b^2 + h^2}}{b^2}\,
\sqrt{E^2 - b^2\left(m^2 + \frac{l^2}{r^2}\right)}\equiv -\frac{h E}{b^2} \pm\frac{(a^2b^2+h^2)\sqrt{R}}{b^2r^2}.
\label{eq:pr_explicit}
\end{equation}
The two signs correspond to the outgoing and incoming branches of the radial motion. For the angular part, we obtain 
\begin{equation}
p_\theta = \pm \sqrt{l^2 - \frac{l_z^2}{\sin^2\theta}}.
\label{eq:ptheta_explicit}
\end{equation}

Thus the abbreviated action \eqref{eq:action} can be written as
\begin{equation}
S = -E t + l_z \phi + \int p_r\, dr + \int p_\theta\, d\theta,
\label{eq:S_separated}
\end{equation}
where the radial and angular integrals are taken along the orbit. The action is a function of the coordinates \((t,\phi,r,\theta)\) and of the constants \((E,l_z,l,m)\). Because the system is completely integrable, we can perform a canonical transformation to new variables \((Q^\mu,P_\mu)\) where the new momenta are precisely these constants:
\begin{equation}
P_0 = m,\quad P_1 = E,\quad P_2 = l_z,\quad P_3 = l.
\label{eq:new_momenta}
\end{equation}
The new coordinates (angle variables) are defined by
\begin{equation}
Q^\mu = \frac{\partial S}{\partial P_\mu},\qquad \mu = 0,1,2,3.
\label{eq:angle_def}
\end{equation}
Using the previously defined quantity \(R\) (see Eq.~\eqref{eq:R_definition}) to simplify the radial integrals, the angle variables are expressed as follows
\begin{align}
Q^0 &= \frac{\partial S}{\partial m} = -m\int \frac{ r^2}{ \sqrt{R}}\, dr,
\label{eq:Q0}\\[4pt]
Q^1 &= \frac{\partial S}{\partial E} = -t + \int\!\left( -\frac{h}{b^2} \pm \frac{E r^2}{b^2 \sqrt{R}} \right) dr,
\label{eq:Q1}\\[4pt]
Q^2 &= \frac{\partial S}{\partial l_z} = \phi - l_z\int\frac{d\theta}{\sin^2\theta\,
\sqrt{l^2 -\sin^{-2}l_z^2}},
\label{eq:Q2}\\[4pt]
Q^3 &= \frac{\partial S}{\partial l} = -l\int \frac{1}{\sqrt{R}}\, dr +l \int \frac{\,d\theta}{\sqrt{l^2 - \sin^{-2}l_z^2}}.
\label{eq:Q3}
\end{align}
These expressions define the action-angle variables \((Q^\mu,P_\mu)\). 

In these variables the Hamiltonian depends only on the new momenta, namely \(H = \frac{1}{2} P_0^2\). Consequently the Vlasov equation \(\{f,H\}=0\) for the distribution function \(f\) reduces, after the canonical transformation, to
\begin{equation}
\frac{\partial f}{\partial Q^0} = 0,
\label{eq:Vlasov_action_angle}
\end{equation}
since \(H\) does not depend on \(Q^0\). By stationarity and spherical symmetry, \(f\) is also independent of \(Q^1,Q^2,Q^3,P_2\). Thus \(f\) can be taken as an arbitrary function of the conserved quantities alone \cite{Rioseco2017a}
\begin{equation}
f = f(P_0,P_1,P_3).
\label{eq:F_conserved}
\end{equation}
This result is essential for constructing steady-state accretion flows: we may choose a distribution function that describes a thermal equilibrium at infinity and then compute the particle current density and the stress-energy tensor by integrating over momentum space.

\section{Relativistic equilibrium distributions}
We assume that the gas surrounding the black hole is in local thermodynamic equilibrium at infinity and the natural choice for the equilibrium distribution is \cite{Rezzolla2013}
\begin{equation}
f_{\infty} = A\,\frac{1}{\exp[(E - \mu)/k_B T] + \varepsilon},
\label{eq:equilibrium_dist}
\end{equation}
where \(A\) is a normalisation constant, \(k_B\) is Boltzmann's constant, \(T\) is the temperature, and \(\mu\) is the chemical potential. The parameter \(\varepsilon\) distinguishes three different types of quantum statistics
\begin{itemize}
    \item \(\varepsilon = +1\): \textbf{Fermi–Dirac statistics}. This describes a gas of fermions (half-integer spin particles) that obey the Pauli exclusion principle. At most one particle can occupy a given quantum state. Examples include electrons, protons, and neutrons. The distribution is essential for describing degenerate matter such as that found in white dwarfs and neutron stars.
    \item \(\varepsilon = 0\): \textbf{Maxwell–Jüttner statistics} \cite{Juttner1911,Juttner1911-2}. This is the relativistic generalisation of the classical Maxwell–Boltzmann distribution. It is valid when quantum effects are negligible, i.e., when the particle density is low or the temperature is high enough that the occupation numbers are small (\(\exp[(E-\mu)/k_B T] \gg 1\)). 
    \item \(\varepsilon = -1\): \textbf{Bose–Einstein statistics}. This describes a gas of bosons (integer spin particles). In contrast to fermions, any number of bosons can occupy the same quantum state. Photons in a blackbody cavity and certain dark matter candidates are typical examples.
\end{itemize}

To simplify the calculation of observable quantities, we consider a gas composed of identical test particles, all having the same rest mass \(m_0\). It is convenient to incorporate this constraint directly into the distribution function by including a delta function \(\delta(P_0 - m_0)\) that selects only those particles with the correct mass, where the delta function does not affect the validity of the Vlasov equation. The resulting distribution function remains a solution of the Vlasov equation throughout the spacetime. Therefore, we adopt the following full distribution function:
\begin{equation}
f = \delta(P_0 - m_0) \, A\,\frac{1}{\exp[(P_1 - \mu)/k_B T] + \varepsilon},
\label{eq:full_dist}
\end{equation}
where \(P_0 = m\) and \(P_1 = E\). At infinity, integration over the mass variable using the delta function recovers the original equilibrium distribution \(f_{\infty}\). This choice greatly simplifies the phase-space integrals in the computation of the particle current density and the stress-energy tensor, as will be shown in the following sections.

It is important to note that while the functional form \eqref{eq:full_dist} is used everywhere, the physical interpretation of \(T\) and \(\mu\) as the temperature and chemical potential is strictly valid only at infinity. In the vicinity of the black hole, these parameters serve merely as convenient labels that characterise the ensemble; the local thermodynamic equilibrium is not necessarily maintained because the gas is collisionless. Nevertheless, the distribution \eqref{eq:full_dist} provides a steady-state, spherically symmetric solution of the relativistic Vlasov equation that is consistent with the asymptotic thermal bath. This approach has been widely adopted in the literature on relativistic kinetic accretion.

\section{The particle current density and the stress energy-momentum tensor}

The particle current density and the stress-energy-momentum tensor are defined as
\begin{align}
J_\mu &= \int p_\mu f(x,p)\, \mathrm{dvol}_x(p), \label{eq:J_def}\\
T_{\mu\nu} &= \int p_\mu p_\nu f(x,p)\, \mathrm{dvol}_x(p), \label{eq:T_def}
\end{align}
with the invariant phase-space volume element
\begin{equation}
\mathrm{dvol}_x(p) = \sqrt{-\det(g^{\mu\nu})}\; d^4p = \sqrt{-\det(g^{\mu\nu})}dp_t\,dp_r\,dp_\theta\,dp_\phi,
\label{eq:vol_element}
\end{equation}
where \(\sqrt{-\det(g^{\mu\nu})} = \frac{1}{r^2\sin\theta\,\sqrt{a^2b^2+h^2}}\). The covariant momenta in terms of the conserved quantities are
gaven by Eqs. (\ref{eq:pr_explicit}) and (\ref{eq:ptheta_explicit}). We further introduce
\begin{align}
p_\theta &= l\cos\sigma, \label{eq:ptheta}\\
p_\phi &= l\sin\theta\sin\sigma, \label{eq:pphi}
\end{align}
and change variables from \((p_t,p_r,p_\theta,p_\phi)\) to \((m^2,E,l_z,l^2)\) and then to \((E,l,\sigma)\). 
The Jacobian of the first transformation is
\begin{equation}
\frac{\partial(m^2,E,l_z,l^2)}{\partial(p_t,p_r,p_\theta,p_\phi)} = 4(g^{rr}p_r+g^{tr}p_t) p_\theta.
\end{equation}
 The second transformation gives
\begin{equation}
dl_z\,dl^2 = 2l^2\sin\theta\,\cos\sigma\,dl\,d\sigma.
\end{equation}
Combining these with \(\sqrt{-\det(g^{\mu\nu})}\), we obtain after canceling \(\cos\sigma\) the volume element
\begin{equation}
\mathrm{dvol}_x(p) = \frac{m l}{ \sqrt{R} }\,dm \,dE\,dl\,d\sigma.
\label{eq:dvol_final}
\end{equation}

We now perform the \(\sigma\) integrals. Using the identities
\begin{equation}
\int_0^{2\pi} \sin\sigma\, d\sigma =\int_0^{2\pi} \cos\sigma\, d\sigma =  \int_0^{2\pi} \sin\sigma\cos\sigma\, d\sigma = 0,
\label{eq:sin_cos_int}
\end{equation}
and
\begin{equation}
\int_0^{2\pi} \sin^2\sigma\, d\sigma = \int_0^{2\pi} \cos^2\sigma\, d\sigma = \pi,
\label{eq:sin2_cos2_int}
\end{equation}
we find that the following components vanish: \(J_\theta, J_\phi, T_{t\theta}, T_{t\phi}\) (linear in \(\cos\sigma\) or \(\sin\sigma\)) and \(T_{\theta\phi}\) (proportional to \(\sin\sigma\cos\sigma\)).
And for the angular diagonal components we obtain \(T_{\theta\theta} = \frac{T_{\phi\phi}}{\sin^2\theta} \).
The remaining non-zero components after the \(\sigma\) integration are:
\begin{align}
J_t &= -2\pi m_0 A \iint \; \frac{El}{ \sqrt{R_0} }\,  \frac{1}{\exp[(E-\mu)/k_B T] + \varepsilon}dE\,dl,\label{Jt}\\[6pt]
J_r &= 2\pi m_0 A \iint \; \frac{l}{b^2}\left(\pm\frac{a^2b^2+h^2}{r^2}- \frac{hE}{ \sqrt{R_0} } \right) \frac{1}{\exp[(E-\mu)/k_B T] + \varepsilon} dE\,dl,\\[6pt]
T_{tt} &= 2\pi m_0 A \iint \; \frac{E^2l}{ \sqrt{R_0} }\, \frac{1}{\exp[(E-\mu)/k_B T] + \varepsilon}dE\,dl,\\[6pt]
T_{tr} &= -2\pi m_0 A \iint \; \frac{El}{b^2}\left(\pm\frac{a^2b^2+h^2}{r^2}- \frac{hE}{ \sqrt{R_0} } \right) \frac{1}{\exp[(E-\mu)/k_B T] + \varepsilon} dE\,dl,\\[6pt]
T_{rr} &= 2\pi m_0 A \iint \; \frac{l\sqrt{R_0} }{b^4}\left(\pm\frac{a^2b^2+h^2}{r^2}- \frac{hE}{ \sqrt{R_0} } \right)^2 \frac{1}{\exp[(E-\mu)/k_B T] + \varepsilon} dE\,dl,\\[6pt]
T_{\theta\theta} &= \pi m_0 A \iint \; \frac{l^3}{ \sqrt{R_0} }\,  \frac{1}{\exp[(E-\mu)/k_B T] + \varepsilon}dE\,dl,\label{Tthth}
\end{align}
where $R_0$ is the function $R$ in the case of $m=m_0$. For the absorption part, there is only the branch "+"; and for the scarring part, there are both the branches "+" and "-". In the above expression, the $r$ component depends particularly on the choice of coordinates. This explicit coordinate dependence will be analyzed in the following sections.


\section{The limits of integration}
In this section, we will discuss the integral limits of Eqs. (\ref{Jt}) - (\ref{Tthth}). For the Schwarzschild black hole we have \(b^2 = 1-2M/r\) and the identity \(a^2b^2+h^2=1\) holds in all coordinate systems. Specializing the general radial equation \eqref{eq:rdot_squared} to a particle of mass \(m_0\) gives
\begin{equation}
R = (E^2-m_0^2)r^4+2M m_0^2 r^3-l^2(r^2-2Mr). \label{eq:radial_schw}
\end{equation}
There are three categories for those particles incident from infinity: those that fall into the black hole and are absorbed, those that are scattered by the black hole and return to infinity, and—lying between these two scenarios—a critical case in which particles neither enter the black hole nor escape to infinity, but instead asymptotically approach an unstable bound orbit. The critical orbits satisfy $R=0$ and $\partial_r R=0$, Solving them yields the well‑known expression (see e.g. Ref. \cite{Li2026-1})
\begin{equation}
l_c^2(E) = \frac{M^2}{2}\,
\frac{27E^4 - 36E^2m_0^2 + 8m_0^4 + E\,(9E^2-8m_0^2)^{3/2}}{E^2-m_0^2}, \label{eq:lc}
\end{equation}
which determined the critical momentum $l=l_c(E)$. Thus, the quantities $J_{\mu}$ and $T_{\mu\nu}$ are naturally  separate into three parts: the absorbed, the scattered, and the critical parts, which are \cite{Li2026-1}
\begin{align}
J_\mu &= J_\mu^{\text{abs}} + J_\mu^{\text{scat}}+J_\mu^{\text{cri}},\qquad \\
T_{\mu\nu} &= T_{\mu\nu}^{\text{abs}} + T_{\mu\nu}^{\text{scat}}+T_{\mu\nu}^{\text{cri}}.
\end{align}
Particles are captured if \(l < l_c(E)\) and scattered if \(l > l_c(E)\).  However, since the critical part happens exactly at $l=l_c$ and $\int_{l_c}^{l_c}f(l)dl=0$, the critical part does not contribute substantially. The critical orbit also satisfies the parametric relations
\begin{equation}
l_c^2 = \frac{M r_c^2 m_0^2}{r_c - 3M},\qquad E = m_0\,\frac{r_c - 2M}{\sqrt{r_c(r_c-3M)}}, \label{eq:rc_lc}
\end{equation}
where $r_c$ is the unstable circular orbit. For the photon orbit, this value is $r_c=3M$.

For a scattered particle to reach a given radius \(r\), the energy must be at least the value obtained by interpreting \(r\) as the critical radius \(r_c\) in \eqref{eq:rc_lc}:
\begin{equation}
E_{\text{min}}(r) = m_0\,\frac{r - 2M}{\sqrt{r(r-3M)}},\qquad r > 3M. \label{eq:Emin_expr}
\end{equation}
This expression diverges as \(r\to 3M^+\), equals \(m_0\) at \(r=4M\), and approaches \(m_0\) from above as \(r\to\infty\).  Since a particle coming from infinity must satisfy \(E \ge m_0\), the expression is valid only for \(3M < r < 4M\); for \(r > 4M\) the minimal energy is \(m_0\), while for \(r < 3M\) no scattered particle can reach that radius (\(E_{\text{min}}=\infty\)).  Hence
\begin{equation}
E_{\text{min}}(r) = \begin{cases}
\infty, & r < 3M,\\
\displaystyle m_0\,\frac{r - 2M}{\sqrt{r(r-3M)}}, & 3M \le r \le 4M,\\
m_0, & r > 4M.
\end{cases} \label{eq:Emin}
\end{equation}
The maximal angular momentum at given \(E\) and \(r\) follows from \eqref{eq:radial_schw} with \(\dot{r}^2\ge0\):
\begin{equation}
l_{\text{max}}^2(E,r) = r^2\left(\frac{E^2}{1-2M/r} - m_0^2\right). \label{eq:lmax}
\end{equation}

Using the general expressions \eqref{eq:radial_schw} for the particle current and stress‑energy tensor, the absorption and scattering contributions are obtained by restricting the integrals as follows:
\begin{itemize}
\item \textbf{Absorption:} \(E \in [m_0,\infty)\), \(l \in [0,\,l_c(E))\).
\item \textbf{Scattering:} \(E \in [E_{\text{min}}(r),\infty)\), \(l \in (l_c(E),\,l_{\text{max}}(E,r)]\).
\end{itemize}
The total quantities are the sum of absorption and scattering parts.

\section{Coordinate-independent physical quantities}

The particle current density \(J_\mu\) and stress-energy tensor \(T_{\mu\nu}\) depend on the choice of coordinates through the metric functions \(a(r), b(r), h(r)\). However, certain combinations of their components are invariant under coordinate transformations that preserve spherical symmetry and stationarity. In this section we identify these invariants: the eigenvalues of \(T^\mu{}_{\ \nu}\) (giving the energy density and pressures) and the particle number density derived from \(J^\mu\). We also clarify the coordinate independence of accretion rates.  Throughout this section we use the fact that the metric \eqref{eq:metric} satisfies the identity \(a^2b^2+h^2=1\), which holds for all coordinate choices considered in this paper.

\subsection{Diagonalisation of the stress-energy tensor}

The mixed tensor \(T^\mu_{\ \nu} = g^{\mu\rho}T_{\rho\nu}\) has the block-diagonal form
\begin{equation}
T^\mu_{\ \nu} =
\begin{pmatrix}
T^t_{\ t} & T^t_{\ r} & 0 & 0 \\
T^r_{\ t} & T^r_{\ r} & 0 & 0 \\
0 & 0 & T^\theta_{\ \theta} & 0 \\
0 & 0 & 0 & T^\phi_{\ \phi}
\end{pmatrix},
\qquad T^\phi_{\ \phi}=T^\theta_{\ \theta}.
\end{equation}
The eigenvalues of the \((t,r)\) block are
\begin{equation}
\lambda_\pm = \frac{T^t_{\ t}+T^r_{\ r} \pm \sqrt{(T^t_{\ t}-T^r_{\ r})^2 + 4T^t_{\ r}T^r_{\ t}}}{2}.
\end{equation}
The energy density \(\rho>0\) is the negative of the negative eigenvalue, and the radial pressure \(p_{\text{rad}}\) is the positive eigenvalue:
\begin{align}
\rho &= -\lambda_- = \frac{1}{2}\left( -T^t_{\ t} - T^r_{\ r} + \sqrt{(T^t_{\ t}-T^r_{\ r})^2 + 4T^t_{\ r}T^r_{\ t}} \right), \label{eq:rho}\\
p_{\text{rad}} &= \lambda_+ = \frac{1}{2}\left( T^t_{\ t} + T^r_{\ r} + \sqrt{(T^t_{\ t}-T^r_{\ r})^2 + 4T^t_{\ r}T^r_{\ t}} \right). \label{eq:prad}
\end{align}
The tangential pressure is simply
\begin{equation}
p_{\text{tan}} = T^\theta_{\ \theta} = \frac{T_{\theta\theta}}{r^2}.
\end{equation}
These three scalars \(\rho, p_{\text{rad}}, p_{\text{tan}}\) are coordinate-independent, as they are eigenvalues of a tensor.

For a perfect fluid at rest, the stress-energy tensor would be diagonal with \(p_{\text{rad}} = p_{\text{tan}} = p\). In the Vlasov gas, however, one generally finds \(p_{\text{rad}} \neq p_{\text{tan}}\) in the finite radii. This indicates that the Vlasov gas is an anisotropic fluid. Following Letelier \cite{Letelier1980}, the stress-energy tensor can be decomposed into a sum of a perfect fluid (with pressure \(p_{\text{tan}}\)) and a null fluid that carries a radial energy flux
\begin{equation}
T^{\mu\nu} = T_{\text{perfect}}^{\mu\nu} + T_{\text{null}}^{\mu\nu}.
\end{equation}
The null component is responsible for the off-diagonal entries \(T^t_{\ r}\) and \(T^r_{\ t}\). 

\subsection{Particle number density and accretion rates}

The particle number density is defined as
\begin{equation}
n = \sqrt{-g_{\mu\nu} J^\mu J^\nu}.
\end{equation}
This quantity is a scalar and therefore coordinate-independent, even though the individual components \(J^\mu\) and the metric depend on the coordinate choice.

For a steady, spherically symmetric flow, the conservation law \(\nabla_\mu J^\mu = 0\) reduces to
\begin{equation}
\frac{\partial}{\partial r}\left( r^2 J^r \right) = 0,
\end{equation}
where we have used the identity \(a^2b^2+h^2=1\) which implies \(\sqrt{-g}=r^2\sin\theta\). Hence the quantity
\begin{equation}
\dot{n} \equiv -4\pi r^2 J^r
\end{equation}
is constant with respect to \(r\).  Evaluating this constant at the event horizon \(r_+\) gives the particle accretion rate (number of particles crossing the horizon per unit time).  Despite the fact that \(J^r\) itself depends on the coordinate system, the product \(-4\pi r^2 J^r\) is invariant because the transformation properties of \(J^r\) exactly compensate the change in the area element. 

Similarly, the energy accretion rate is defined using the conserved current associated with the timelike Killing vector \(\partial_t\).  The stress-energy conservation \(\nabla_\mu T^{\mu}_{\ t}=0\) leads to
\begin{equation}
\frac{\partial}{\partial r}\left( r^2 T^r_{\ t} \right) = 0,
\end{equation}
so that
\begin{equation}
\dot{E} \equiv -4\pi r^2 T^r_{\ t}
\end{equation}
is also constant and coordinate-independent.  For a gas of identical neutral particles of mass \(m_0\), the mass accretion rate is simply
\begin{equation}
\dot{M} = m_0 \dot{n}.
\end{equation}
These accretion rates are the primary physical quantities that determine how the black hole's mass evolves over time.

In summary, while the components \(J^\mu\) and \(T^{\mu\nu}\) depend on the coordinate system, the derived scalars \(n\), \(\rho\), \(p_{\text{rad}}\), \(p_{\text{tan}}\) and the integrated rates \(\dot{n}\), \(\dot{E}\) are invariant.  The identity \(a^2b^2+h^2=1\) ensures that the formulas for the accretion rates take the simple form \(-4\pi r^2 J^r\) and \(-4\pi r^2 T^r_{\ t}\) in all coordinates considered, while the underlying physical quantities remain unchanged.

\section{Asymptotic behavior at infinity}

In this section we study the asymptotic behavior of the particle current density \(J_\mu\) and the stress-energy tensor \(T_{\mu\nu}\) in the limit \(r\to\infty\).  Only the scattering part contributes; the absorption part vanishes at large distances.  We use the general metric of Sec.~II, which satisfies \(a^2b^2+h^2=1\), and specialize to the Schwarzschild black hole with \(b^2=1-2M/r\).  The radial kinetic term for a particle of mass \(m_0\) is rewritten as
\begin{equation}
R_0 = r^4\dot{r}^2 = X - Y l^2,\qquad 
X = r^4\bigl(E^2 - b^2 m_0^2\bigr),\quad Y = r^2 b^2. \label{eq:R0_def}
\end{equation}
For \(r\to\infty\) the integration domain simplifies to
\[
E\in[m_0,\infty),\qquad l\in[l_c(E),\, l_{\max}(E,r)],\quad l_{\max}=\sqrt{X/Y}.
\]
The critical angular momentum \(l_c(E)\) is independent of \(r\) and given by Eq.~\eqref{eq:lc}. We also define \(R_{0c}=X-Y l_c^2\).

We now expand the relevant quantities in powers of \(1/r\).  For the Schwarzschild black hole,
\begin{align}
b^2(r) &= 1 - \frac{2M}{r},\\
h(r) &= h_0 + \frac{h_1}{r} + \cdots,\\
a^2(r) &= 1 - h_0^2 + \frac{2M(1-h_0^2)-2h_0h_1}{r} + \cdots,
\end{align}
with \(h_0\) and \(h_1\) depending on the coordinate choice.  Defining \(\Delta = E^2 - m_0^2\), we obtain the following integrals:
\begin{align}
I_1 &\equiv \int_{l_c}^{l_{\max}} l\, dl = \frac12\left(\frac{X}{Y} - l_c^2\right)=r^2\left(\frac{\Delta}{2}+\frac{M}{r}(m_0^2+\Delta) +\mathcal{O}\!\left(\frac{1}{r^2}\right)\right), \label{eq:I1}\\
I_2 &\equiv\int_{l_c}^{l_{\max}} \frac{l}{\sqrt{R_0}}\, dl = \frac{\sqrt{R_{0c}}}{Y}=\sqrt{\Delta}+\frac{M}{r}\frac{m_0^2+2\Delta}{\sqrt{\Delta}}+ \mathcal{O}\!\left(\frac{1}{r^2}\right), \label{eq:I2}\\
I_3 &\equiv \int_{l_c}^{l_{\max}} \frac{l^3}{\sqrt{R_0}}\, dl = \frac{\sqrt{R_{0c}}}{3Y^2}\bigl(2A + B l_c^2\bigr)=\frac{2r^2}{3}\left( \Delta^{3/2}+\frac{M\sqrt{\Delta}}{r}(3m_0^2+4\Delta)+\mathcal{O}\!\left(\frac{1}{r^2}\right)  \right), \label{eq:I3}\\
I_4 &\equiv \int_{l_c}^{l_{\max}} l\sqrt{R_0}\, dl = \frac{R_{0c}^{3/2}}{3Y}=\frac{r^4}{3}\left( \Delta^{3/2}+\frac{M\sqrt{\Delta}}{r}(3m_0^2+2\Delta)+\mathcal{O}\!\left(\frac{1}{r^2}\right)  \right) \label{eq:I4}.
\end{align}
Inserting these expansions into the integrals \eqref{Jt} – \eqref{Tthth} and defining the average \(\langle Q(E)\rangle = \int_{m_0}^\infty  \frac{Q(E)}{\exp[(E-\mu)/k_BT]+\varepsilon}\, dE\), we obtain the asymptotic results

\paragraph{Coordinate‑independent components:}
\begin{align}
J_t^{\infty} &= -4\pi A m_0 \left[ \langle \sqrt{\Delta} E \rangle + \frac{M}{r}\left\langle E\left(\frac{m_0^2}{\sqrt{\Delta}} + 2\sqrt{\Delta}\right) \right\rangle +\mathcal{O}\!\left(\frac{1}{r^2}\right)\right], \label{eq:Jt_asym}\\
T_{tt}^{\infty} &= 4\pi A m_0 \left[ \langle \sqrt{\Delta} E^2 \rangle + \frac{M}{r}\left\langle E^2\left(\frac{m_0^2}{\sqrt{\Delta}} + 2\sqrt{\Delta}\right) \right\rangle +\mathcal{O}\!\left(\frac{1}{r^2}\right)\right], \label{eq:Ttt_asym}\\
T_{\theta\theta}^{\infty} &= \frac{4\pi r^2}{3} A m_0  \left[ \langle \Delta^{3/2} \rangle + \frac{M}{r}\left\langle \sqrt{\Delta}(3m_0^2+4\Delta)\right\rangle +\mathcal{O}\!\left(\frac{1}{r^2}\right)\right]. \label{eq:Ttheta_asym}
\end{align}

\paragraph{Coordinate‑dependent components:}
\begin{align}
J_r ^{\infty}&= -4\pi A m_0 \Bigl(h_0 \langle E\sqrt{\Delta} \rangle  + \frac{1}{r} \langle \frac{E }{\sqrt{\Delta}} (h_0M(m_0^2+4\Delta)+h_1\Delta)\rangle + \mathcal{O}\!\left(\frac{1}{r^2}\right)\Bigr), \label{eq:Jr_asym}\\[6pt]
T_{tr} ^{\infty}&= 4\pi A m_0 \Bigl(h_0 \langle E^2\sqrt{\Delta} \rangle  + \frac{1}{r} \langle \frac{E^2 }{\sqrt{\Delta}} (h_0M(m_0^2+4\Delta)+h_1\Delta)\rangle + \mathcal{O}\!\left(\frac{1}{r^2}\right)\Bigr), \label{eq:Ttr_asym}\\[6pt]
T_{rr}^{\infty} &= \frac{4}{3}\pi A m_0 \Bigl( \langle \Delta^{3/2} \rangle + 3h_0^2 \langle E^2\sqrt{\Delta} \rangle \nonumber\\\quad &+\frac{3}{r}\langle \frac{M\Delta(m_0^2+2\Delta)+E^2h_0(Mh_0(m_0^2+6\Delta)+2h_1\Delta) }{\sqrt{\Delta}} \rangle+ \mathcal{O}\!\left(\frac{1}{r^2}\right)\Bigr) . \label{eq:Trr_asym}
\end{align}

We now compute the particle number density $n_{\infty} = \sqrt{-g_{\mu\nu} J^\mu_{\infty} J^\nu_{\infty}}$.
Using the contravariant components \(J^\mu = g^{\mu\nu} J_\nu\) and the leading constant parts of \(J_t\) and \(J_r\) from Eqs.~\eqref{eq:Jt_asym} and \eqref{eq:Jr_asym}, together with the inverse metric components to order \(r^0\),
\[
g^{tt}=-(1-h_0^2),\quad g^{tr}=-h_0,\quad g^{rr}=1,
\]
we obtain
\begin{align}
n_{\infty}  =\sqrt{ -g^{tt}(J_t)^2 - 2g^{tr}J_tJ_r - g^{rr}(J_r)^2}=4\pi A m_0\langle \sqrt{\Delta}E\rangle.\label{eq:n_asym}
\end{align}
All dependence on \(h_0\) and \(h_1\) has disappeared, confirming that the particle number density is a coordinate scalar.

We then compute the mixed tensor \(T^{\infty}{}^\mu_{\ \nu}=g^{\mu\rho}T^{\infty}_{\rho\nu}\) to leading order.  Using the constant parts of the covariant components, we find
\begin{align}
T^{\infty}{}^t_{\ t} &= g^{tt}T^{\infty}_{tt}+g^{tr}T^{\infty}_{rt} = -4\pi A m_0 \langle\sqrt{\Delta}E^2\rangle,\\
T^{\infty}{}^r_{\ r} &= g^{rr}T^{\infty}_{rr}+g^{tr}T^{\infty}_{tr} = \frac{4}{3}\pi A m_0 \langle\Delta^{3/2}\rangle,\\
T^{\infty}{}^t_{\ r} &= g^{tt}T^{\infty}_{tr}+g^{tr}T^{\infty}_{rr}
= -4\pi A m_0 h_0 \left(\langle E^2\sqrt{\Delta}\rangle + \frac13\langle\Delta^{3/2}\rangle\right),\\
T^{\infty}{}^r_{\ t} &= g^{rr}T^{\infty}_{rt}+g^{tr}T^{\infty}_{tt} = 0.
\end{align}
Thus the \((t,r)\)-block of \(T^{\infty}{}^\mu_{\ \nu}\) is upper triangular; the energy density \(\rho^{\infty} = -T^{\infty}{}^t_{\ t}\) and the radial pressure \(p^{\infty}_{\text{rad}} = T^{\infty}{}^r_{\ r}\) are therefore
\begin{align}
\rho^{\infty} &= 4\pi A m_0 \langle\sqrt{\Delta}E^2\rangle, \label{eq:rho_asym}\\
p ^{\infty}_{\text{rad}} &= \frac{4}{3}\pi A m_0 \langle\Delta^{3/2}\rangle. \label{eq:prad_asym}
\end{align}
The tangential pressure is obtained from the angular component:
\begin{equation}
p^{\infty}_{\text{tan}} =\frac{T^{\infty}_{\theta\theta}}{r^2} = \frac{4}{3}\pi A m_0 \langle\Delta^{3/2}\rangle = p^{\infty}_{\text{rad}}. \label{eq:ptan_asym}
\end{equation}
Hence at infinity the Vlasov gas is isotropic, and the pressure satisfies the equation of state of a relativistic ideal gas.  Moreover, all expressions are manifestly independent of \(h_0\) and \(h_1\), confirming their coordinate invariance.

\section{Behavior near the event horizon}

We now investigate the behavior of the particle current density and stress-energy tensor in the vicinity of the event horizon $r = 2M$. It is convenient to introduce the small parameter
\begin{equation}
x \equiv r - 2M,
\end{equation}
and expand the metric functions $b^2(r)$, $h(r)$ and $a^2(r)$ in powers of $x$.  Using the identity $a^2 b^2 + h^2 = 1$ we obtain
\begin{align}
b^2(r) &= \frac{x}{2M} - \frac{x^2}{4M^2} + \mathcal{O}(x^3), \label{eq:b2exp}\\
h(r) &= H_0 + H_1 x + \mathcal{O}(x^2), \label{eq:hexp}\\
a^2(r) &= \frac{2M(1-H_0^2)}{x} + 1 - H_0^2 - 4H_0 H_1 M + \mathcal{O}(x), \label{eq:a2exp}
\end{align}
where $H_0$ and $H_1$ are constants that parametrize the coordinate choice.  
For the Schwarzschild black hole the auxiliary functions $X$ and $Y$ defined in Eq.~(\ref{eq:R0_def}) become
\begin{equation}
X = 16M^4 E^2 + 8M^3(4E^2 - m_0^2)\, x + \mathcal{O}(x^2), \qquad
Y = 2M x + x^2 + \mathcal{O}(x^3). \label{eq:ABexp}
\end{equation}

Near the event horizon \(r=2M\) only the absorption part contributes.   
Hence the integration interval for the angular momentum is $l \in [0, l_c(E)]$, where $l_c(E)$ is given by Eq.~(\ref{eq:lc}) and does not depend on $x$. The radial momentum corresponds to the incoming branch (minus sign in $p_r$). The integrals $I_1$--$I_4$ appearing in Eqs.~(\ref{eq:I1})--(\ref{eq:I4}) can be evaluated in closed form for the limits $[0,l_c]$.  
Expanding the results in powers of $x$ gives
\begin{align}
\mathcal{ I}_1 &\equiv \int_{0}^{l_{c}} l\, dl= \frac{l_c^2}{2}, \label{eq1:I1}\\[4pt]
\mathcal{ I}_2 &\equiv\int_{0}^{l_{c}} \frac{l}{\sqrt{R_0}}\, dl=\frac{\sqrt{X}-\sqrt{R_{0c}}}{Y}= \frac{l_c^2}{8M^2E} 
      + \frac{l_c^2\left[l_c^2+8M^2(m_0^2-4E^2)\right]}{256M^5E^3}x + \mathcal{O}(x^2), \label{eq1:I2}\\[4pt]
\mathcal{ I}_3 &\equiv\int_{0}^{l_{c}} \frac{l^3}{\sqrt{R_0}}\, dl=\frac{2X^{3/2}-\sqrt{R_{0c}}(2X+Y l_c^2)}{3Y^2}= \frac{l_c^4}{16M^2E} 
      + \frac{l_c^4\left[l_c^2+6M^2(m_0^2-4E^2)\right]}{384M^5E^3}x + \mathcal{O}(x^2), \label{eq1:I3}\\[4pt]
\mathcal{ I}_4 &\equiv \int_{0}^{l_{c}} l\sqrt{R_0}\, dl=\frac{X^{3/2}-R_{0c}^{3/2}}{3Y}= 2M^2E l_c^2 
      +\frac{l_c^2\left[l_c^2+8M^2(m_0^2-4E^2)\right]}{16E M}x + \mathcal{O}(x^2). \label{eq1:I4}
\end{align}

Substituting these expansions into the general expressions (\ref{Jt})-(\ref{Tthth}) for the absorption part, we obtain the following near-horizon expansions for the covariant components of the particle current and the stress-energy tensor:
\begin{align}
J_t^{H} &= -\frac{\pi m_0 A}{4M^2}\left(\langle l_c^2\rangle
     + \left\langle \frac{l_c^2  \left(l_c^2+8 M^2 \left(m_0^2-4 E^2\right)\right)}{32 E^2 M^3}\right\rangle x
      +\mathcal{O}(x^2) \right), \label{eq:JtNH}\\[8pt]
J_r^{H} &= \frac{\pi m_0 A}{2M x}(1-H_0)\bigl(\langle l_c^2\rangle -\left\langle   \frac{l_c^2 \left(16 E^2M^2(1+2H_1M)+H_0M^2(l_c^2+8M^2(m_0^2-2E^2))\right)}{32 E^2 (1-H_0) M^3}\right\rangle x\nonumber\\
&+ \mathcal{O}(x^2)\bigr), \label{eq:JrNH}\\[8pt]
T_{tt}^{H} &=  \frac{\pi m_0 A}{4M^2}\left(\langle E l_c^2\rangle
     + \left\langle \frac{l_c^2  \left(l_c^2+8 M^2 \left(m_0^2-4 E^2\right)\right)}{32 E M^3}\right\rangle x
      +\mathcal{O}(x^2) \right), \label{eq:TttNH}\\[8pt]
T_{tr}^{H} &= -\frac{\pi m_0 A}{2M x}(1-H_0)\bigl(\langle E l_c^2\rangle -\left\langle   \frac{l_c^2 \left(16 E^2M^2(1+2H_1M)+H_0M^2(l_c^2+8M^2(m_0^2-2E^2))\right)}{32 E (1-H_0) M^3}\right\rangle x\nonumber\\
&+ \mathcal{O}(x^2)\bigr), \label{eq:TtrNH}\\[8pt]
T_{rr}^{H} &= \frac{\pi m_0 A(1-H_0)^2}{x^2}\bigl( \langle E l_c^2\rangle-\left\langle \frac{l_c^2 \left(8 M^2 \left(8 E^2 H_1 M+(H_0+1) m_0^2\right)+(H_0+1) l_c^2\right)}{32 E (1-H_0) M^3}\right\rangle x \nonumber\\& +\mathcal{O}(x^2)\bigr), \label{eq:TrrNH}\\[8pt]
T_{\theta\theta}^{H} &= \frac{\pi m_0 A}{16M^2}\left( \left\langle \frac{l_c^4}{E}\right\rangle +     \left\langle \frac{l_c^4 \left(6 M^2 \left(m_0^2-4 E^2\right)+l_c^2\right)}{24 E^3 M^3}\right\rangle x
                  +\mathcal{O}(x^2) \right). \label{eq:TthetataNH}
\end{align}

Although the individual components $J_r^{H}$, $T_{tr}^{H}$ and $T_{rr}^{H}$ diverge as $x^{-1}$ or $x^{-2}$, physical observables constructed from them remain finite at the horizon.  
The particle number density $n_{H} = \sqrt{-g_{\mu\nu}J^\mu_H J^\nu_H}$ is found to be
\begin{equation}
n_H = \frac{\pi m_0 A}{8\sqrt{2}M^3}\,\sqrt{\langle l_c^2\rangle\langle \frac{l_c^2(l_c^2+8m_0^2M^2)}{E^2}\rangle }+ \mathcal{O}(x),
\label{eq:nNH}
\end{equation}
where the leading term is independent of $H_0$ and $H_1$.  
Diagonalizing the mixed stress-energy tensor $T^\mu_{\ \nu}$ on the horizon gives the energy density and pressures, given to
\begin{align}
\rho^H & = \frac{\pi m_0 A}{192M^4}\,   \left(3 f_2+2\sqrt{3}\sqrt{f_1 f_3}\right) + \mathcal{O}(x), \label{eq:rhoNH}\\
p^H_{\rm rad} & = \frac{\pi m_0 A}{192M^4}\,   \left(-3 f_2+2\sqrt{3}\sqrt{f_1 f_3}\right) + \mathcal{O}(x), \label{eq:pradNH}\\
p^H_{\rm tan} &= T^\theta_{\ \theta} = \frac{\pi m_0 A}{64 M^4}\,\left\langle \frac{l_c^4}{E}\right\rangle + \mathcal{O}(x), \label{eq:ptanNH}
\end{align}
where 
\begin{align}
    f_1=\langle E l_c^2\rangle,\quad
    f_2=\langle \frac{l_c^2(l_c^2+8m_0^2M^2)}{E}\rangle,\quad
    f_3=\langle \frac{l_c^2(l_c^4+12l_c^2m_0^2M^2+48m_0^4M^4)}{E^3}\rangle.
\end{align}
All three quantities are manifestly coordinate independent at leading order.

Finally, evaluating the accretion rates $\dot n = -4\pi r^2 J^r$ and $\dot E = -4\pi r^2 T_t^{\ r}$ at the horizon, we obtain the exact constants
\begin{align}
\dot n &= 4\pi^2 m_0 A\,\langle l_c^2\rangle, \label{eq:ndotNH}\\
\dot E &= 4\pi^2 m_0 A\,\langle E l_c^2\rangle, \label{eq:EdotNH}
\end{align}
which are free of any coordinate artifacts and coincide with the values obtained by integrating the absorption part at any radius.

The above results confirm that, despite the strong coordinate dependence of the individual tensor components near the horizon, the physical properties of the accreting Vlasov gas --- density, pressure, and accretion rates --- are well defined and invariant.

\section{The asymptotic results of Maxwell--J\"uttner distribution and the accretion rate of thermodynamic entropy}

In this section we specialize the general expressions obtained in the preceding sections to the Maxwell--J\"uttner distribution, which describes a classical relativistic gas in thermal equilibrium.  For this case the statistical parameter $\varepsilon$ in the distribution function (\ref{eq:full_dist}) is set to zero and the chemical potential $\mu$ is taken to vanish, so that $f=A\delta(P_0-m_0)e^{- \epsilon z}$, where we have introduce the dimensionless energy $\epsilon=\frac{E}{m_0}$ and the dimensionless inverse temperature $z \equiv \frac{m_0}{k_{\mathrm{B}}T} $.

At large distances the geometry is flat and the gas is isotropic.  The leading contributions to the particle current density and the stress-energy tensor are given by the scattering part, whose asymptotic expressions have been derived in Sec. VIII. The following integrals are useful
\begin{align}
  \langle\sqrt{\Delta}E\rangle &= m_0^3\int_1^{\infty} \sqrt{\epsilon^2-1}\; \epsilon\; \mathrm{e}^{-z\epsilon}\,d\epsilon=m_0^3\frac{K_2(z)}{z},\\
    \langle\sqrt{\Delta}E^2\rangle &= m_0^4\int_1^{\infty} \sqrt{\epsilon^2-1}\; \epsilon^2\; \mathrm{e}^{-z\epsilon}\,d\epsilon=m_0^4\left(\frac{K_1(z)}{z}+3\frac{K_2(z)}{z^2}\right),\\
    \langle \Delta^{3/2}\rangle&=m_0^4\int_1^{\infty} (\epsilon^2-1)^{3/2}\; \mathrm{e}^{-z\epsilon}\,d\epsilon=m_0^4\cdot 3\frac{K_2(z)}{z^2},
\end{align}
where $K_{1}(z)$ and $K_{2}(z)$ are the Bessel functions of the first kind and second kind. Using these integrals, one can easily obtain
\begin{align}
n_{\infty} &= 4\pi A m_0^4 \frac{K_2(z)}{z}, \label{eq:n_inf}\\
\rho_{\infty} &= 4\pi A m_0^5 \left( \frac{K_1(z)}{z} + 3\frac{K_2(z)}{z^2} \right), \label{eq:rho_inf}\\
p_{\infty} &= 4\pi A m_0^5 \frac{K_2(z)}{z^2}. \label{eq:p_inf}
\end{align}
One verifies that the ideal gas equation of state $p_{\infty}=n_{\infty}k_{\mathrm{B}}T$ holds identically, which shows the gas behaves exactly as an isotropic perfect fluid at infinity.  The average energy of a classical particle at temperature $T$ is
\begin{equation}
    \bar{E}=\frac{\rho_{\infty}}{n_{\infty}}=m_0\left(\frac{K_1(z)}{K_2(z)}+\frac{3}{z}\right).
\end{equation}
In the non-relativistic limit $z\rightarrow \infty $, this value approaches
\begin{equation}\label{MBEn}
    \bar{E}=m_0\left( 1+\frac{3}{2z} \right)=m_0+\frac{3}{2}k_BT,
\end{equation}
where we have used the limits $\lim\limits_{z\rightarrow \infty}\frac{K_1(z)}{K_2(z)}\approx 1-\frac{3}{2}z$. The first term in the above expression \eqref{MBEn} denotes the rest energy $m_0c^2$ (here \(c=1\)), and the second term $\frac{3}{2}k_BT$ corresponds to the translational kinetic energy arising from the temperature $T$, which is none other than the classical Maxwell–Boltzmann result.

\begin{figure*}
\centering
\setlength{\tabcolsep}{2pt} 
  \renewcommand{\arraystretch}{1} 
  \begin{tabular}{cc} 
  \includegraphics[width=0.45\textwidth]{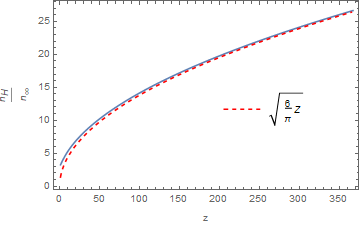} &
    \includegraphics[width=0.45\textwidth]{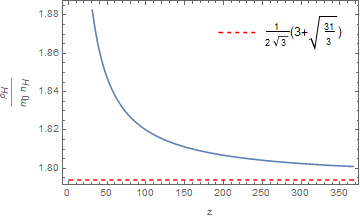} \\
    \includegraphics[width=0.45\textwidth]{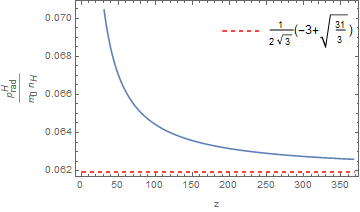} &
    \includegraphics[width=0.45\textwidth]{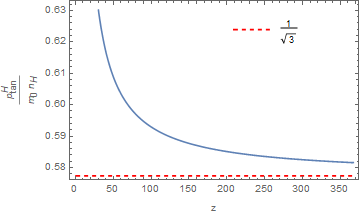}
  \end{tabular}
\caption{The numerical values of $\frac{n_{H}}{n_{\infty}},\frac{\rho_H}{n_H}, \frac{p_{\text{rad}}^H}{n_H}, \frac{p_{\text{rad}}^H}{n_H} $ for the finite $z$.}
\label{Fig:numeri}
\end{figure*}
On the horizon the absorption part dominates and the relevant physical quantities are given by the expressions (\ref{eq:nNH})--(\ref{eq:ptanNH}) together with the definitions of $f_1$, $f_2$, $f_3$. They are complicated expressions. However, similer to the reference \cite{Rioseco2017a}, one can obtain simpler relevant expressions in the low-temperature limit $z\rightarrow \infty $. Let's introduce the dimensionless variables $y = z(\epsilon-1)$. Then $\epsilon = 1 + y/z$, $\mathrm{d}\epsilon = \mathrm{d}y/z$, and the exponential factor becomes $\mathrm{e}^{-z\epsilon}= \mathrm{e}^{-z}\mathrm{e}^{-y}$.  Any average $\langle Q(E)\rangle = \int_{m_0}^{\infty} Q(E) \mathrm{e}^{-E/k_{\mathrm{B}}T} dE$ transforms to
\[
\langle Q\rangle = m_0 \mathrm{e}^{-z} \int_0^{\infty} Q\!\left(m_0\left(1+\frac{y}{z}\right)\right) \mathrm{e}^{-y}\,\frac{\mathrm{d}y}{z}.
\]
One can obtain the following expressions in the limit $z\rightarrow \infty $
\begin{align}
    l_c^2&\approx \frac{E l_c^2}{m_0} \approx 16 m_0^2 M^2,\\
    \frac{l_c^2(l_c^2+8m_0^2M^2)}{E^2}&\approx m_0\frac{l_c^2(l_c^2+8m_0^2M^2)}{E} \approx 384 m_0^2 M^4,\\
    \frac{l_c^2(l_c^4+12 l_c^2 m_0^2M^2+48m_0^4M^4)}{E^3}&\approx 7936 m_0^3 M^6,\\
    \frac{l_c^4}{E^2}&\approx 256 m_0^3 M^4.
\end{align}
Since the leading terms of these expressions are constant, the averages of these quantities are proportional to $\frac{e^{-z}}{z}$, for example
\begin{equation}
    \langle l_c^2\rangle \approx m_0 \frac{e^{-z}}{z}\int_{0}^{\infty} 16m_0^2 M^2 e^{-y}dy=16m_0^3 M^2\frac{e^{-z}}{z}.
\end{equation}
After some similar mathematical calculations, we obtain
\begin{align}
n_H &= 4\sqrt{3}\pi m_0^4 A\frac{e^{-z}}{z},\\
\rho_H &= \frac{2}{3}(9+\sqrt{93})\pi m_0^5 A\frac{e^{-z}}{z},\\
p_{\text{rad}}^{H} &= \frac{2}{3}(-9+\sqrt{93})\pi m_0^5 A\frac{e^{-z}}{z},\\
p_{\text{tan}}^{H} &= 4\pi m_0^5 A\frac{e^{-z}}{z}.
\end{align}
The ratios with the particle density $n_H$ are therefore finite in the limit $z\to\infty$
\begin{align}
\frac{\rho_H}{n_H} &= \frac{m_0}{2\sqrt{3}}\left(3+\sqrt{\frac{31}{3}}\right),\\
 \frac{p_{\text{rad}}^H}{n_H} &= \frac{m_0}{2\sqrt{3}}\left(-3+\sqrt{\frac{31}{3}}\right),\\
\frac{p_{\text{tan}}^H}{n_H} &= \frac{m_0}{\sqrt{3}},
\end{align}
and compression ratio $\frac{n_{H}}{n_{\infty}}$ is
\begin{equation}
    \lim_{z\to\infty}\frac{1}{\sqrt{z}} \frac{n_{H}}{n_{\infty}}=\lim_{z\to\infty} \frac{\sqrt{3}}{\sqrt{z}}\frac{e^{-z}}{K_2(z)}=\sqrt{\frac{6}{\pi}}.
\end{equation}
For finite values of $z$, numerical integration can be employed, and the results are shown in the Fig.\ref{Fig:numeri}. The accretion rates are
\begin{equation}
    \dot{n}\approx \frac{\dot{E}}{m_0} \approx 64\pi^2m_0^5A M^2\frac{e^{-z}}{z},\\
\end{equation}
and the mass accretion rate $\dot{M}=m_0 \dot{n}$ approximate to
\begin{equation}
    \lim_{z\to\infty}\frac{1}{\sqrt{z}} \frac{\dot{M}}{n_{\infty}}=\lim_{z\to\infty} \frac{16\pi m_0 M^2}{\sqrt{z}}\frac{e^{-z}}{K_2(z)}=16\sqrt{2\pi}m_0M^2.
\end{equation}
These asymptotic results coincide exactly with those obtained by Rioseco and Sarbach \cite{Rioseco2017a} in their low-temperature analysis, thereby confirming the consistency of our horizon expansions. 

We can actually compute this more precisely. The average energy of a single particle falling into the black hole is defined as
\begin{equation}
    \bar{E}=\frac{\dot{E}}{\dot{n}}=\frac{\langle El_c^2\rangle}{\langle l_c^2\rangle}.
\end{equation}
In the classical limit $z\rightarrow \infty $, expanding to first order in $1/z$, we obtain
\begin{equation}
    \bar{E}\approx \frac{1+\frac{5}{z}}{1+\frac{4}{z}}\approx m_0\left( 1+\frac{1}{z}\right)=m_0+k_BT,
\end{equation}
where the first term denotes the rest energy, and the second term denotes the translational kinetic energy arising from the temperature $T$, which is smaller than $\frac{3}{2}k_BT$. This implies that the particles captured by the black hole are subject to angular momentum selection. Particles with higher energies are more prone to possessing large angular momenta, which renders them more easily scattered back to infinity by the centrifugal barrier. Hence, the energy spectrum of the captured particles is cut off and softened, and their mean energy is indeed below the overall average.

Moreover, we shall now consider the entropy accretion rate. The entropy flux through a spherical surface of arbitrary radius can be defined as
\begin{equation}
    \dot{S}=4\pi r^2k_B\int_{abs} f(x,p)\ln f(x,p)~ p^r \, \mathrm{dvol}_x(p).
\end{equation}
Inserting the expression of $f=A\delta(P_0-m_0)e^{- \frac{E}{k_BT}}$, one obtains 
\begin{align}
      \dot{S}&=-k_B\ln A \bigg(-4\pi \int_{abs} f(x,p)p^r \, \mathrm{dvol}_x(p) \bigg)+\frac{1}{ T}\bigg(-4\pi \int_{abs} Ef(x,p)p^r \, \mathrm{dvol}_x(p) \bigg) \nonumber \\
      &=-k_B\ln A ~\dot{n}+\frac{\dot{E}}{T}.
\end{align}
We define the specific entropy as $\sigma_{abs}=\frac{\dot{S}}{\dot{n}}$, where
\begin{equation}
    \sigma_{abs}=-k_B\ln A +\frac{\bar{E}}{T}=k_B(-\ln A+z+1).
\end{equation}
At infinity, the system is described by a Maxwell–Jüttner distribution, and the value of the specific entropy is given by $\sigma=k_B(-\ln A +z+\frac{5}{2})$. This indicates that the entropy per particle of the accreted particles is lower than the global average by $\frac{3}{2}k_B$. 

\section{numerical results}

\subsection{Particle current density and particle density}
\begin{figure*}
\centering
\setlength{\tabcolsep}{2pt} 
  \renewcommand{\arraystretch}{1} 
  \begin{tabular}{cc} 
  \includegraphics[width=0.45\textwidth]{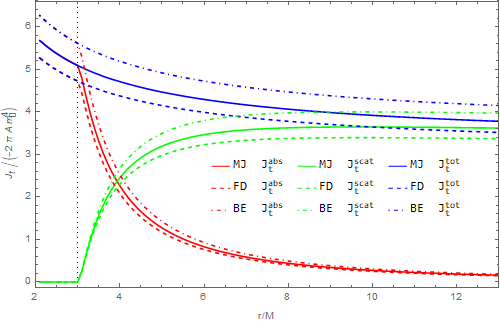} &
    \includegraphics[width=0.45\textwidth]{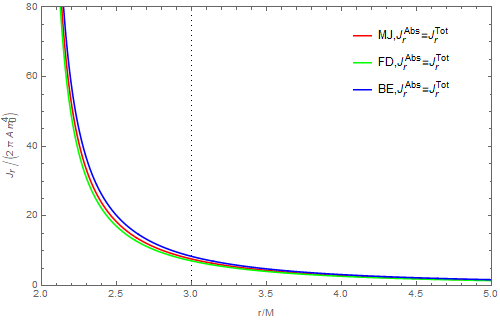} \\
    \includegraphics[width=0.45\textwidth]{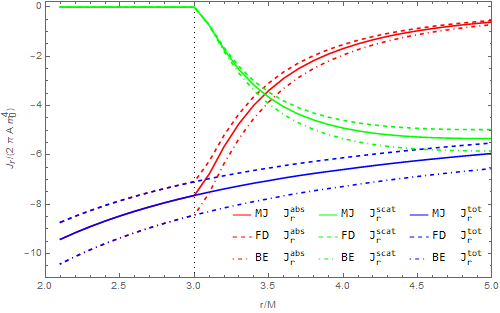} &
    \includegraphics[width=0.45\textwidth]{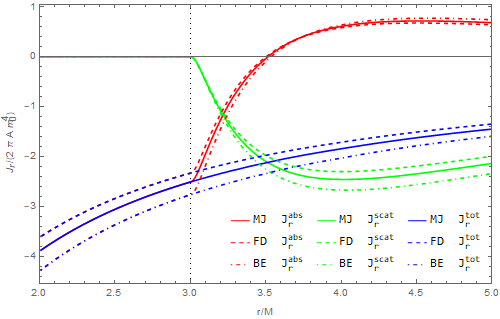}
  \end{tabular}
\caption{The numerical values of $J_t$ and $J_r$ for $z=1$. The metric used for the image on the right side of the first row is $h=0$. At this point, since the scattered part of the incident branch and the outgoing branch cancel each other out, only the absorbing part remains. The image on the left side of the second row uses the metric $h=1$, and the part on the right side of the second row uses the metric $h=2/r$.}
\label{Fig:JtJr}
\end{figure*}
\begin{figure*}
\centering
\setlength{\tabcolsep}{2pt} 
  \renewcommand{\arraystretch}{1} 
  \includegraphics[width=0.5\textwidth]{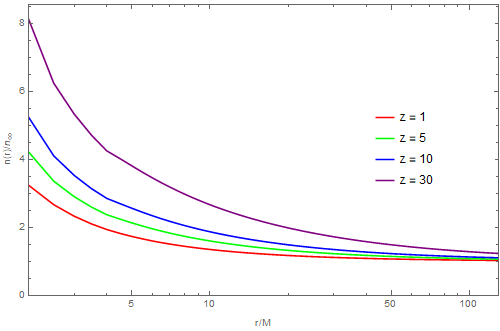} 
\caption{The behavior of the compression rate $n(r)/n_{\infty}$ as a function $r/M$ for different temperatures in Maxwell-J\"uttner statistic.}
\label{Fig:nratio}
\end{figure*}

We perform the numerical calculation using Wolfram Mathematica. Without loss of generality, we choose $M=1$ and $\mu=0$. Fig. \ref{Fig:JtJr} shows the numerical results of $J_t$ and $J_r$, where $J_r$ depends on the metric form and is calculated by three different metric. The three different coordinate systems exhibit different asymptotic behaviors. When $h = 0$, $J_r$ is divergent at horizon and approaches zero at infinity. When $h = 1$, both $J_r$ at horizon and at infinity are finite values given by Eq. (\ref{eq:JrNH}) and Eq. (\ref{eq:Jr_asym}). While, the particle density are coordinate independent, see Fig. \ref{Fig:nratio}.

\begin{table}[h]
    \centering
    \begin{tabular}{p{4.5cm} p{2.5cm} p{2.5cm} p{2.5cm} p{2.5cm}}
        \toprule
        distribution  & z=1 & z=5 & z=10 & z=30 \\
        \midrule
        \(\varepsilon = 0\) & 3.24405 & 4.22358 & 5.24868 & 8.13214 \\
        \midrule
        \(\varepsilon = 1\)   & 3.22868 & 4.22151 & 5.24866 & 8.13214 \\
        \midrule
        \(\varepsilon = -1\)  & 3.26997 & 4.22567 & 5.24871 & 8.13214 \\
        \bottomrule
    \end{tabular}
 \caption{The numerical results of $\frac{n_H}{n_{\infty}}$ for three different statistics. The chemical potential is chosen to be $\mu=0$.}
 \label{tab:numofnratio1}
\end{table}

The numerical results of $\frac{n_H}{n_{\infty}}$ for Maxwell-J\"uttner distribution are shown in Fig. \ref{Fig:nratio}. The numerical results $\frac{n_H}{n_{\infty}}$ for the other two distributions are also very close, as can be seen from the table \ref{tab:numofnratio1}. The reasons are as follows: the integration has the expression
\begin{equation}
     \langle\sqrt{\Delta}E\rangle = m_0^3\int_1^{\infty} \sqrt{\epsilon^2-1}\; \epsilon\; \frac{1}{\mathrm{e}^{z\epsilon}+\varepsilon}\,d\epsilon,
\end{equation}
where the influence of the parameter $\varepsilon$ on the integral is very limited, and this effect becomes even smaller as $z$ increases. For this reason, it is difficult to distinguish different distributions from the graph of $\frac{n_H}{n_{\infty}}$.

\subsection{Energy density, radial and tangential pressures }

\begin{figure*}
\centering
\setlength{\tabcolsep}{2pt} 
  \renewcommand{\arraystretch}{1} 
  \begin{tabular}{cc} 
  \includegraphics[width=0.48\textwidth]{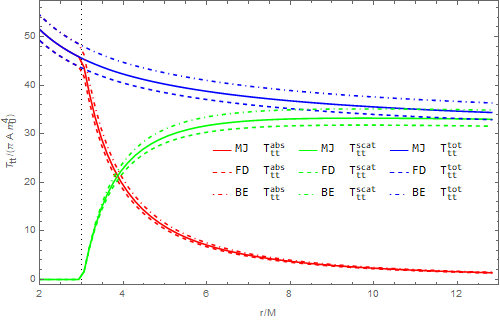} &
    \includegraphics[width=0.48\textwidth]{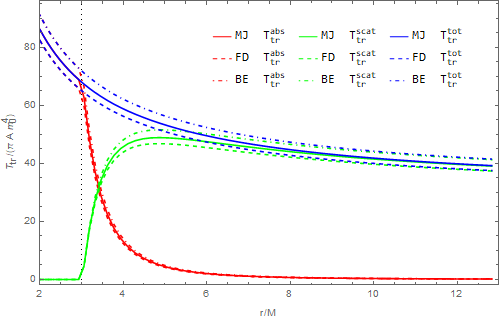} \\
    \includegraphics[width=0.48\textwidth]{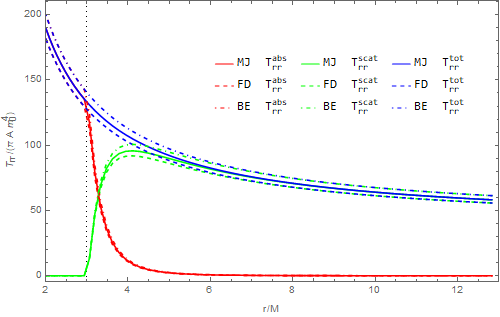} &
  \end{tabular}
\caption{The numerical values of $T_{tt}, T_{tr}$ and $T_{rr}$ for $z=1$. We use the metric $h=1$.}
\label{Fig:TttTrr}
\end{figure*}

\begin{figure*}
\centering
\setlength{\tabcolsep}{2pt} 
  \renewcommand{\arraystretch}{1} 
  \begin{tabular}{cc} 
  \includegraphics[width=0.48\textwidth]{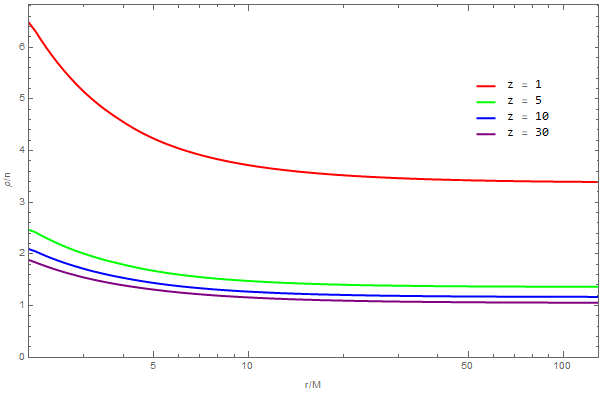} &
    \includegraphics[width=0.48\textwidth]{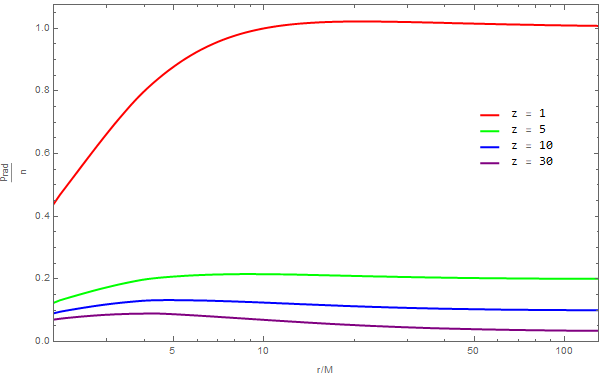} \\
    \includegraphics[width=0.48\textwidth]{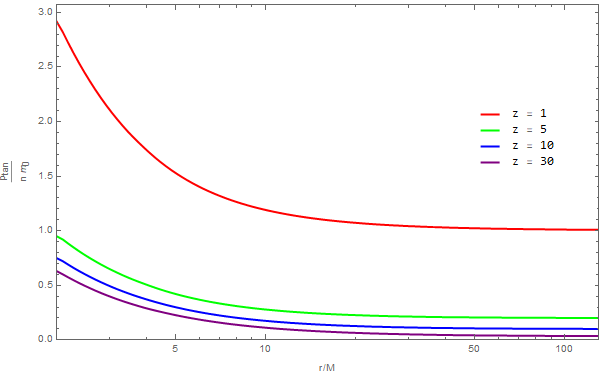} & \includegraphics[width=0.48\textwidth]{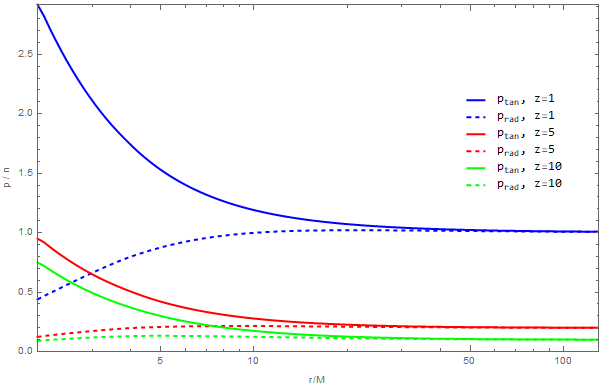}
  \end{tabular}
\caption{The numerical values of $\frac{\rho}{n m_0}, \frac{p_{\text{rad}}}{n m_0}$ and $\frac{p_{\text{tan}}}{n m_0}$ as the function of $r/M$ for Maxwell-J\"{u}ttner distribution.}
\label{Fig:rho_ptan_prad}
\end{figure*}

The energy-momentum tensor $T_{\mu\nu} (\mu,\nu=t,r)$ are plotted in Fig. \ref{Fig:TttTrr}. By diagonalizing $T^{\mu}{}_{\nu}$, the coordinate-independent physical quantities $\frac{\rho}{n}, \frac{p_{\text{rad}}}{n}$ and $\frac{p_{\text{tan}}}{n}$ for Maxwell-J\"{u}ttner distribution are plotted in Fig \ref{Fig:rho_ptan_prad}. These figures are consistent with Ref. \cite{Rioseco2017b}, showing that $p_{\text{rad}}=p_{\text{tan}}$ at infinity; However, as $r/M$ decreases, $p_{\text{rad}}$ gradually becomes smaller than
$p_{\text{tan}}$.

In Table \ref{tab:rho_p}, we present the numerical results $\frac{\rho}{n m_0}, \frac{p_{\text{rad}}}{n m_0}$ and $\frac{p_{\text{tan}}}{n m_0}$ for the asymptotic near horizon and asymptotic infinity in three different distributions. They show that the differences between these results are very small. Therefore, the numerical results $\frac{\rho}{n m_0}, \frac{p_{\text{rad}}}{n m_0}$ and $\frac{p_{\text{tan}}}{n m_0}$  of the other two distributions will be very similar to those in Figure \ref{Fig:rho_ptan_prad}, and we will not draw them separately here. The results in Table \ref{tab:rho_p} were also used to examine the asymptotic behavior in Figure \ref{Fig:rho_ptan_prad}.

\begin{table}
\centering
\scriptsize
\setlength{\tabcolsep}{2.5pt}
\begin{tabular}{p{1cm} p{1.5cm} p{1.5cm} p{1.5cm} p{1.5cm}p{1.5cm} p{1.5cm} p{1.5cm} p{1.5cm} p{1.5cm}}
\toprule
 &  & \multicolumn{2}{c}{\(z=1\)} & \multicolumn{2}{c}{\(z=5\)} & \multicolumn{2}{c}{\(z=10\)} & \multicolumn{2}{c}{\(z=30\)} \\

\cmidrule(lr){3-4} \cmidrule(lr){5-6} \cmidrule(lr){7-8} \cmidrule(lr){9-10}
& & \(r/M=2\) & \(r/M=\infty\) &\(r/M=2\) &\(r/M=\infty\) &\(r/M=2\) &\(r/M=\infty\) & \(r/M=2\)&\(r/M=\infty\)
\\
\toprule
 & \(\varepsilon=0 \) &6.48658 & 3.37044 &2.4693 & 1.36185 &2.09571 & 1.16699 & 1.88471 & 1.05202 \\
\cmidrule{2-10}
\(\frac{\rho}{n m_0} \) &\(\varepsilon=1 \)  &6.68286 & 3.46204 &2.47024 & 1.36224 & 2.09571& 1.16699 &1.88471 & 1.05202 \\
\cmidrule{2-10}
 & \(\varepsilon=-1 \)  & 6.22378 & 3.25017 &2.46836 & 1.36145 &2.09571 & 1.16699 & 1.88471& 1.05202\\
\cmidrule{1-10}
 & \(\varepsilon=0 \) &0.439051 & 1&0.124059 &0.2 & 0.0906328&   0.1&0.0706736 & 0.0333333\\
\cmidrule{2-10}
\(\frac{p_{\text{rad}}}{n m_0} \) &\(\varepsilon=1 \)  &0.454003 & 1.03457&0.124141 & 0.200202 & 0.0906331& 0.100001 &0.0706736 & 0.0333333 \\
\cmidrule{2-10}
 & \(\varepsilon=-1 \) & 0.418946 & 0.954133 &0.123977 & 0.199797 & 0.0906325& 0.0999993 &0.0706736 &0.0333333 \\
\cmidrule{1-10}
  & \(\varepsilon=0 \) & 2.91865& 1 &0.950425 &0.2 &0.749912 & 0.1 &0.630925 & 0.0333333\\
\cmidrule{2-10}
\(\frac{p_{\text{tan}}}{n m_0} \) &\(\varepsilon=1 \)  &3.01316 & 1.03457 &0.950929 & 0.200202 & 0.749914& 0.100001 &0.630925 & 0.0333333 \\
\cmidrule{2-10}
 & \(\varepsilon=-1 \) & 2.79163 & 0.954133 &0.949919 & 0.199797 &0.74991 &  0.0999993 & 0.630925 & 0.0333333\\
\bottomrule
 \end{tabular}
  \caption{The asymptotic numerical results of $\frac{\rho}{n m_0}, \frac{p_{\text{rad}}}{n m_0}$ and $\frac{p_{\text{tan}}}{n m_0}$ for three different statistics. }
 \label{tab:rho_p}
\end{table}

\section{conclusion and discussion}
In this work, we have systematically investigated the coordinate dependence of the stationary, spherically symmetric accretion of a relativistic Vlasov gas onto a Schwarzschild black hole. By working within the most general stationary metric that allows for off-diagonal $g_{tr}$ components, we have derived the equations of motion, constructed action-angle variables, and obtained exact integral expressions for the particle current density and the stress-energy tensor. Our analysis covers three distinct statistical distributions---Fermi-Dirac, Maxwell-J\"uttner, and Bose-Einstein---and we have performed both asymptotic expansions (at infinity and near the horizon) and full numerical integrations at finite radii. The central result is that while the individual covariant components of $J_\mu$ and $T_{\mu\nu}$ depend explicitly on the chosen coordinate system, all physically observable quantities---namely, the particle number density $n$, energy density $\rho$, radial and tangential pressures $p_{\mathrm{rad}}$ and $p_{\mathrm{tan}}$, as well as the accretion rates $\dot{n}$ and $\dot{E}$---are manifestly coordinate-invariant. This invariance is achieved through appropriate combinations of tensor components and is guaranteed by the identity $a^2 b^2 + h^2 = 1$ that holds for all coordinate representations considered.

A key physical insight emerging from our study is the role of angular momentum in filtering the accreted particles. At infinity, the gas is isotropic and obeys the classical ideal gas equation of state, with mean energy $m_0 + \frac{3}{2} k_B T$. However, particles that are captured by the black hole have, on average, lower energy: in the non-relativistic limit, their mean energy is $m_0 + k_B T$. This reduction arises because high-energy particles tend to carry large angular momenta, which makes them more likely to be scattered away by the centrifugal barrier rather than being absorbed. Consequently, the specific entropy of the accreted particles is lower than the global average by $\frac{3}{2}k_B$, indicating that the accretion process effectively selects the lower-entropy, lower-energy portion of the phase-space distribution. These findings highlight that black hole accretion in the kinetic regime is not merely a passive capture process but involves a dynamical filtering mechanism that modifies the thermodynamic properties of the infalling matter.

From a methodological standpoint, our formulation does not rely on any particular coordinate system, thus providing a unified framework that can be readily extended to other black hole spacetimes, such as Reissner-Nordstr\"om or Kerr, and even to modified theories of gravity with asymptotically flat solutions. The use of action-angle variables and the explicit construction of the invariant phase-space volume element ensure that the Vlasov equation is solved consistently, and that the resulting distribution function remains a steady-state solution throughout the spacetime. The asymptotic expansions at the horizon and at infinity, together with the numerical results for finite radii, confirm the robustness of our conclusions across all regimes. We have also shown that the differences among the three quantum statistics become negligible at low temperatures, as expected, since the classical Maxwell-J\"uttner limit is recovered.

In summary, our work establishes that the accretion of a collisionless kinetic gas onto a Schwarzschild black hole can be described in a fully coordinate-independent manner, with unambiguous definitions for all physical observables. The results not only resolve potential ambiguities arising from different metric representations but also provide a solid foundation for future investigations of more complex accretion scenarios, including rotating black holes, charged black holes, and systems with external perturbations. The entropy reduction and energy filtering effects we have identified may have important implications for the thermodynamics of black hole growth and the evolution of particle distributions in strong gravitational fields.

\textbf{Acknowledgement}
This work was supported in part by the National Natural Science Foundation of China (Grant No. 12505071) and by the Research Foundation of the Education Bureau of Hunan Province, China (Grant No. 25B0635). It was also supported in part by the Key Laboratory of Information Detection and Intelligent Processing Technology of the Hunan Provincial Department of Education, and by the Applied Characteristic Subject of Hunan Province, "Electronic Science and Technology".


\end{document}